\documentclass[final]{svjour3}
\usepackage{graphicx}
\usepackage{amssymb}
\usepackage[numbers]{natbib}
\makeatletter
\journalname{Journal of Low Temperature Physics}

\bibpunct{}{}{,}{s}{}{,}

\begin{document}

\newcommand{\hdblarrow}{H\makebox[0.9ex][l]{$\downdownarrows$}-}
\title{Quantum Dynamics of Atom-molecule BECs in a Double-Well Potential}

\author{A. Motohashi \and T. Nikuni}

\institute{Department of Physics, Tokyo University of Science,\\ 1-3 Kagurazaka, Shinjuku-ku, Tokyo 162-8601, Japan\\
\email{j1208710@ed.kagu.tus.ac.jp}}
\date{June.15.2009}

\maketitle

\keywords{Bose-Einstein condensation,Bose-Josephson Junction}

\begin{abstract}

We investigate the dynamics of two-component Bose-Josephson junction composed of atom-molecule BECs. Within the semiclassical approximation, the multi-degree of freedom of this system permits chaotic dynamics, which does not occur in single-component Bose-Josephson junctions. By investigating the level statistics of the energy spectra using the exact diagonalization method, we evaluate whether the dynamics of the system is periodic or non-periodic within the semiclassical approximation. Additionally, we compare the semiclassical and full-quantum dynamics.  

\end{abstract}

\section{Introduction}
Since its experimental realization, Bose-Einstein condensation in dilute atomic gases has been offering opportunities to research macroscopic quantum phenomena. Especially, one of the most fascinating macroscopic quantum phenomena is the Josephson effect between two Bose-Einstein condensations(BECs) trapped in a double-well potential, which is called the Bose-Josephson Junction(BJJ). BJJ has been realized experimentally, which allowed direct observation of the dynamics of macroscopic wavefunctions directly\cite{BJJ}. Though in BJJ the spatial coherence of BECs is focussed, the Josephson effect occurs not only between spatial separated BECs but also between internal degrees of freedom in a single BEC. In particular, the Josephson-like effects between atomic and molecular states have been discussed theoretically\cite{Vardi}\cite{Santos}\cite{Duncan}. In this paper, we consider atom-molecule BECs in a double-well potential. In this two component BJJ, the internal atom-molecule tunneling has significant influence on the ground state property and dynamics. As shown later, this internal tunneling induces the localized ground states. As for the dynamics, the multi-degree of freedom of this system permits chaotic dynamics, by adding the atom-molecule tunneling as a new degree of freedom. 

First, we explain the semiclassical dynamics of BJJ. In a single component BJJ, the dynamical degrees of freedom are the relative particle number and the relative phase, so the dimension of the system is two. Since the trajectories in the phase space cannot intersect with each other, significantly complex trajectories cannot exist in 2D systems. However, in higher dimensional systems, the trajectories in the phase space travel around much more freely than 2D dynamical systems. This leads to richer dynamical behaviors in a two component BJJ, and the chaotic dynamics can occur. 

In general full-quantum systems, although the chaotic dynamics can not occur(see Ref. 5 and references there in), the quantum signature of the semiclassical non-periodicity appears in the level spacing distribution\cite{Shudo_Saito}. In our system, we can control the multiplicity of degrees of freedom by varying the atom-molecule internal tunneling strength. Based on these considerations, we investigate the statistical property of energy spectra and compare the semiclassical and full-quantum dynamics.

\section{Model and approximations}
In this section, we explain our model and approximations. The second-quantized Hamiltonian for Bose atoms and molecules can be written as
\begin{eqnarray}
&&\hat{H} = \sum_{i=a,b} \int d \mathbf{r} \Bigg[ \frac{ \hbar^2 }{ 2m_{i} } \nabla \hat{\Psi} _{i}^{ \dag }  \nabla \hat{\Psi}_{i} + V_{ \rm{ext} } ( \mathbf{ r }  )  \hat{ \Psi }_{i}^{ \dag } \hat{ \Psi }_{i} \Bigg] + \frac{g_{i}}{2} \sum_{ i=a,b } \int d \mathbf{r} \hat{ \Psi }_{i}^{ \dag } \hat{ \Psi }_{i}^{ \dag } \hat{ \Psi }_{i} \hat{ \Psi }_{i} \nonumber \\
 && \qquad + g_{ab} \int d \mathbf{r} \hat{ \Psi }_{a}^{ \dag } \hat{ \Psi }_{b}^{ \dag } \hat{ \Psi }_{b} \hat{ \Psi }_{a} - \lambda \int d \mathbf{r} \left( \hat{ \Psi }_{b}^{ \dag } \hat{ \Psi }_{a} \hat{ \Psi }_{a} + \hat{ \Psi }_{b}  \hat{ \Psi }_{a}^{\dag}  \hat{ \Psi }_{a}^{\dag} \right) + \delta \int d \mathbf{r} \hat{ \Psi }_{b}^{ \dag } \hat{ \Psi }_{b} ,  \qquad 
\end{eqnarray}
where $\hat{ \Psi }_{a}$ and $\hat{ \Psi }_{b}$ represent  field operators for Bose atoms and molecules respectively, $\lambda$ is the coupling strength between atomic and molecular states, $\delta$ is the energy difference between atoms and molecules, and $V_{ \rm{ext} } ( \mathbf{ r } )$ is the double-well potential. The inter-atomic, the inter-molecule, and the atom-molecule interactions can be approximated in terms of the s-wave scattering lengths as $g_{i} = 4 \pi \hbar^2 a_{s i} / m_{i}, g_{ab} = 6 \pi \hbar^2 a_{s ab} / m_{a}({\it i}=a,b ,m_{b}=2m_{a})$. Here, $m_{a}$ is the mass of a Bose atom. Furthermore, we introduce the four-mode approximation, which concentrates on condensate modes only, and neglect the effect of the particles occupying other modes. Then, field operators can be approximated as $\hat{ \Psi }_{a} \simeq \Phi_{aL} \hat{a}_{L} + \Phi_{aR} \hat{a}_{R}, \hat{ \Psi }_{b} \simeq \Phi_{bL} \hat{b}_{L} + \Phi_{bR} \hat{b}_{R}$, where $\Phi_{aL}, \Phi_{aR}$($\Phi_{bL}, \Phi_{bR}$) are the wavefunctions of the atomic(molecular) condensate modes in the left well and the right well respectively. $\hat{a}_{L}, \hat{a}_{R}$($\hat{b}_{L}, \hat{b}_{R}$) are annihilation operators for the atomic(molecular) condensate modes in the left well and the right well respectively. Applying these approximations to Eq.(1), we obtain the quantum four-mode Hamiltonian (four-mode model). 
\begin{eqnarray}
\hat{H} &=& - J_{a} \big( a_{L}^{\dag} a_{R} +  a_{R}^{\dag} a_{L} \big) - J_{b} \big( b_{L}^{\dag} b_{R} + b_{R}^{\dag} b_{L} \big) + \Delta \big( b_{L}^{\dag} b_{L} + b_{R}^{\dag} b_{R} \big)  \nonumber \\
&& + \frac{ U_{a} }{2} \big( a_{L}^{\dag} a_{L}^{\dag} a_{L} a_{L} + a_{R}^{\dag} a_{R}^{\dag} a_{R} a_{R} \big) + \frac{ U_{b} }{2} \big( b_{L}^{\dag} b_{L}^{\dag} b_{L} b_{L} + b_{R}^{\dag} b_{R}^{\dag} b_{R} b_{R} \big) \nonumber \\
&& + U_{ab} \big( a_{L}^{\dag} a_{L} b_{L}^{\dag} b_{L} + a_{R}^{\dag} a_{R} b_{R}^{\dag} b_{R} \big) \nonumber \\
&& - g \big( b_{L}^{\dag} a_{L} a_{L} + b_{R}^{\dag} a_{R} a_{R} + b_{L} a_{L}^{\dag} a_{L}^{\dag} + b_{R} a_{R}^{\dag} a_{R}^{\dag} \big) . \quad
\label{4_q_H}
\end{eqnarray}
Here, the parameters are defined as follows :
\begin{eqnarray}
&& J_{i} \equiv - \int d \mathbf{r} \Phi_{iL}^{*} \left[ - \frac{\hbar^2}{ 2m_{i} } \nabla^{2} + V_{ \rm{ext} } ( \mathbf{r} ) \right] \Phi_{iR} ,
\label{eq:p_1} \\
&& E_{i}^{0} \equiv  \int d \mathbf{r} \Phi_{iL}^{*} \left[ - \frac{\hbar^2}{ 2m_{i} } \nabla^{2} + V_{ \rm{ext} } ( \mathbf{r} ) \right] \Phi_{iL}  =  \int d \mathbf{r} \Phi_{iR}^{*} \left[ - \frac{\hbar^2}{ 2m_{i} } \nabla^{2} + V_{ \rm{ext} } ( \mathbf{r} ) \right] \Phi_{iR} , \quad
\label{eq:p_3} \\
&& U_{i} \equiv g_{i} \int d \mathbf{r} | \Phi_{iL} |^{4} = g_{i} \int d \mathbf{r} | \Phi_{iR} |^{4} ,
\label{eq:p_4} \\
&& U_{ab} \equiv g_{ab} \int d \mathbf{r} | \Phi_{aR} |^{2} | \Phi_{bR} |^{2} = g_{ab} \int d \mathbf{r} | \Phi_{aL} |^{2} | \Phi_{bL} |^{2} ,
\label{eq:p_6}
 \\
&& g \equiv \lambda \int d \mathbf{r} \Phi_{bL}^{*} \Phi_{aL} \Phi_{aL} = \lambda \int d \mathbf{r} \Phi_{bR}^{*} \Phi_{aR} \Phi_{aR} ,
\label{eq:p_7}
 \\
&& \Delta \equiv \delta \int d \mathbf{r} | \Phi_{bL} |^{2} + E_{b}^{0} - 2 E_{a}^{0} = \delta \int d \mathbf{r} | \Phi_{bR} |^{2} + E_{b}^{0} - 2 E_{a}^{0} ,
\label{eq:p_9}
\end{eqnarray}
where $i = a$ represents atomic BEC modes, $i = b$ represents molecular BEC modes, and L,R express the left well and right well respectively. And also, we use the notation as $z_{a} = \left( a_{L}^{\dag} a_{L} - a_{R}^{\dag} a_{R} \right) / N$ as needed.

\section{Results}
First, we investigate the effect of the atom-molecule tunneling on the ground state. Next, we will show the relation between the level statistics and the periodicity of dynamics. We choose the parameters as follows. The ratio $\Lambda = N U_{a} / ( 2 J_{a} )$ is estimated as $15$ in the single-component experiment\cite{BJJ}, and the atomic interaction strength normalized by the atomic tunneling strength can be obtained as $U_{a} / J_{a} \simeq 3 \times 10^{-2}$. We use this value for the atomic interaction strength. As for the molecular interaction strength, we suppose that the molecular scattering length is the same as the atomic one and that the shape of condensate wavefunctions of atoms and molecules are also the same. Under this condition $U_{b} = U_{a} / 2$ from Eq. (\ref{eq:p_4}). In this study we set the total particle number as ${\it N} = 20$. We next consider the atom-molecule interaction. The experiment\cite{Heinzen_Molecule} indicates that the atom-molecule interaction of $^{87} \textrm{Rb}$ is attractive, and thus we choose as $U_{ab} / J_{a} \simeq - 2.3 \times 10^{-2}$. Finally, we set $\Delta / J_{a} = - 1$.

\subsection{Phase transition of the ground state}

Within the semiclassical approximation, the particle localization transition in the ground state is induced by the atom-molecule internal tunneling\cite{semi}. The ground state at $\sqrt{N} g / J_{a} < 2.55$ is symmetric, which has equal particle populations in each well. At $\sqrt{N} g / J_{a} \simeq 2.55$, the ground state become degenerate. For $\sqrt{N} g / J_{a} > 2.55$, the ground state obtains non-equal particle population. Here, we investigate this transition within a full-quantum treatment. By the exact diagonalization of the four-mode Hamiltonian (\ref{4_q_H}), we obtain the eigenstates and energy eigenvalues. 

\begin{figure}
\begin{minipage}{14pc}
\includegraphics[%
  width=0.8\linewidth,
  keepaspectratio]{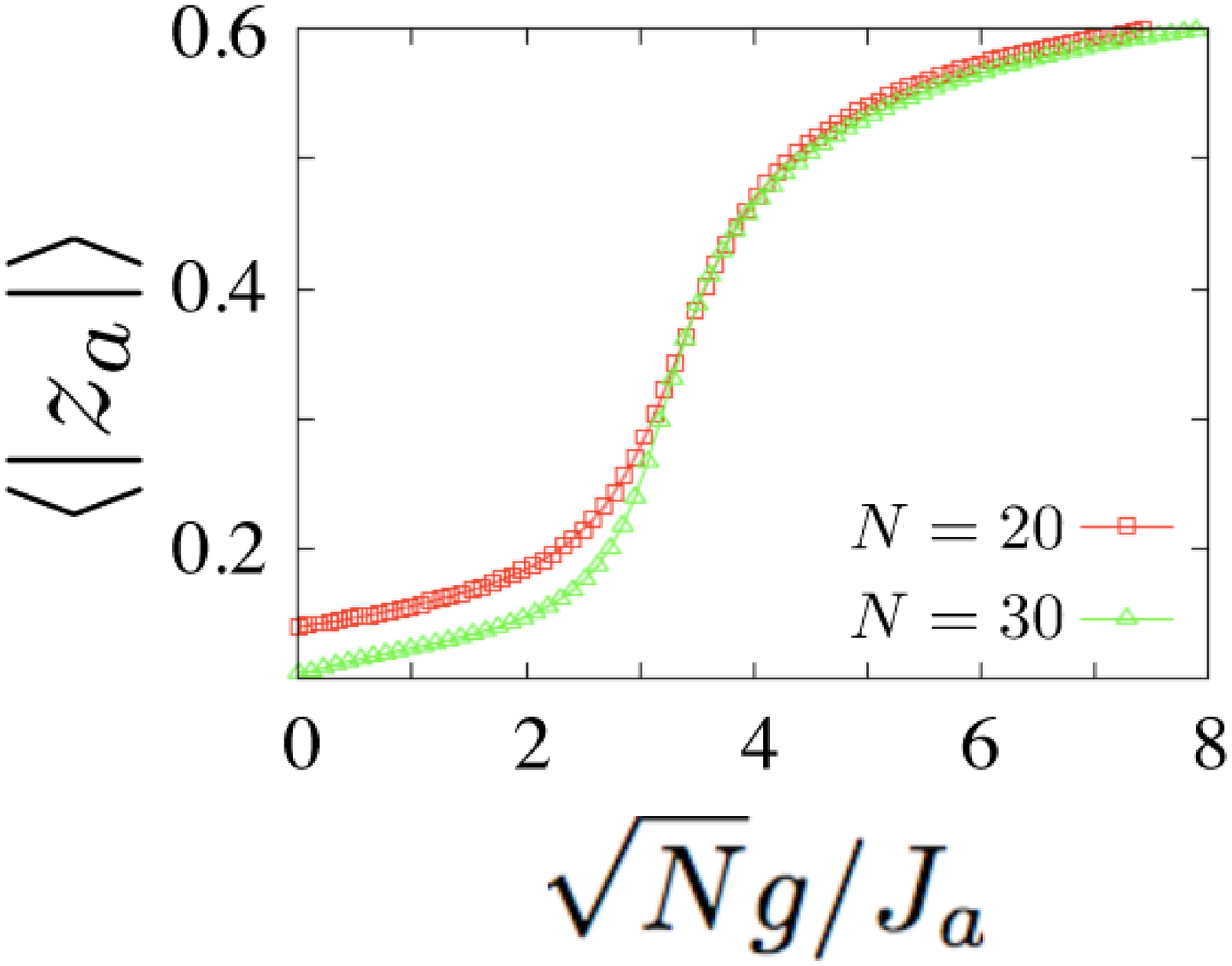}
\caption{(color online)The $g$-dependence of $\langle | z_{a} | \rangle$.}
\label{rel_za}
\end{minipage}\hspace{0.5pc}
\begin{minipage}{14pc}
\includegraphics[%
  width=0.8\linewidth,
  keepaspectratio]{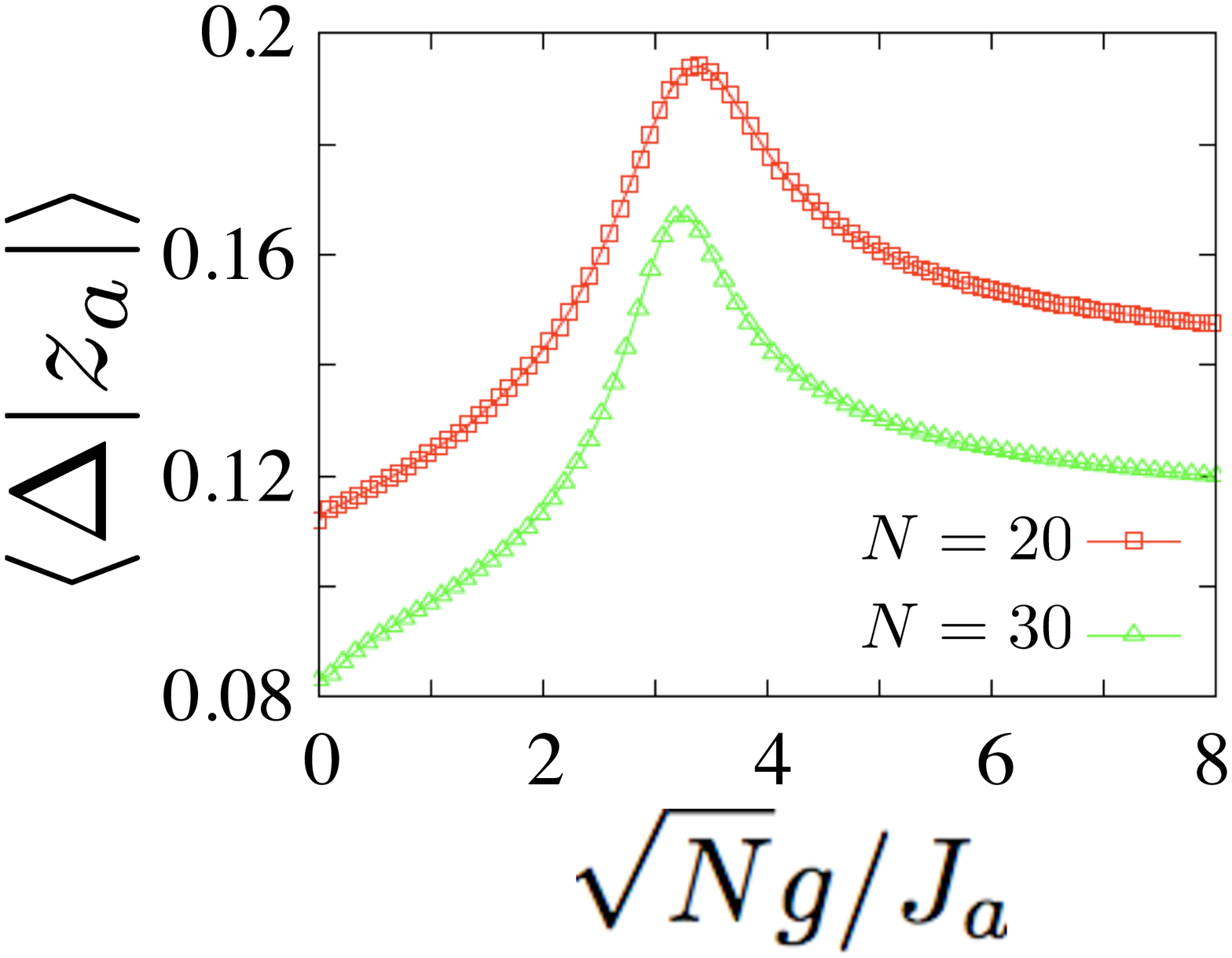}
\caption{(color online)$\langle \Delta | z_{a} | \rangle$ of $N = 20$(red points) and $N = 30$(green points). $\langle \Delta | z_{a} | \rangle$ is defined as $\langle \Delta | z_{a} | \rangle \equiv \sqrt{ \langle z_{a}^2 \rangle - \langle | z_{a} | \rangle^2 }$.}
\label{fluc}
\end{minipage}\hspace{0.5pc}
\end{figure}

\begin{figure}
\begin{center}
\includegraphics[%
  width=0.6\linewidth,
  keepaspectratio]{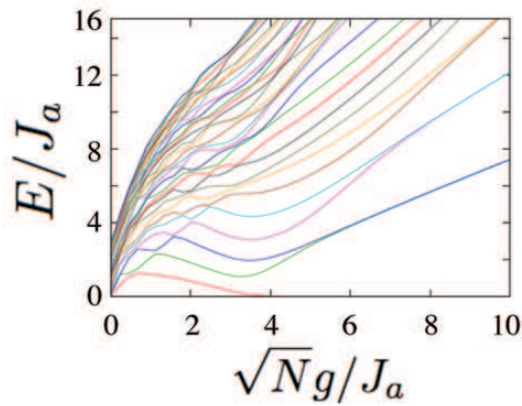}
\caption{(color online)$g$-dependence of the energy spectra normalized by $J_{a}$. Lower 30 levels are presented.}
\label{N_20_spectra_30pon}
\end{center}
\end{figure}

The $g$-dependence of $\langle | z_{a} | \rangle$ is presented in Fig. \ref{rel_za}. $ \langle | z_{a} | \rangle$ increases rapidly at $\sqrt{ N } g / J_{a} \simeq 3.5$. From Fig. \ref{fluc}, we see that the fluctuations of $| z_{a} |$ is maximum at this point. Furthermore, the ground state become degenerate for $\sqrt{ N } g / J_{a} \ge 3.5$, as shown below. The $g$-dependence of excitation energies is shown in Fig. \ref{N_20_spectra_30pon}. Increasing the atom-molecule tunneling $g$, the first excitation energy goes to zero at $\sqrt{N} g / J_{a} \simeq 3.5$. For larger $g$, the ground state and some low lying energy levels become degenerate. These facts indicate that a phase transition to the localized ground state occurs at $\sqrt{N} g / J_{a} \simeq 3.5$. 
Finally, we comment on the system-size dependence of the results. We find that the results for $\langle | z_{a} | \rangle$ are qualitatively the same for different values of $N$. The size dependence appears more prominently in fluctuations; as shown in Fig. \ref{fluc}, the peaks of $\langle \Delta | z_{a} | \rangle$ at the critical point are sharpened for larger $N$. This indicates the occurrence of the quantum phase transition in the thermodynamic limit.

\subsection{Level statistics and dynamics}

Within a semiclassical approximation, we found the existence of non-periodic dynamics\cite{JPCS}. This motivates us to investigate the full-quantum counterpart. 

In this section, we investigate the effect of the atom-molecule tunneling on the periodicity of dynamics. The semiclassical dynamics is periodic in the small and large $\sqrt{N} g / J_{a}$ region, however, non-periodic in the intermediate region. Considering this by full-quantum treatment, we show how the level spacing distributions change by varying the strength of the atom-molecule tunneling. From level spacing distribution, one can decide whether the dynamics of the system is periodic or non-periodic\cite{Shudo_Saito}. 

When a system exhibits regular motions in a semiclassical approximation, level spacing distribution shows Poisson-type distribution $P ( S ) = e^{-S}$ except some cases\cite{Berry_Tabor}, where $S$ is the level spacing normalized by mean level spacing. On the other hand, when the semiclassical dynamics of the system is chaotic, level spacing distribution is Wigner-type distribution $P ( S ) = \left( \pi / 2 \right) S \exp{ ( - \pi S^2 / 4 ) }$. In this section, we compare the level spacing distribution of the four-mode Hamiltonian (\ref{4_q_H}) with Wigner and Poisson distribution in order to investigate the periodicity of semiclassical dynamics. In addition, we show that the time evolutions in a semiclassical approximation and a full-quantum treatment\cite{Tonel} are different in a chaotic regime. Before the analysis, we need to classify the energy spectra according to the symmetry of the system\cite{Chaos_OL}. Since the double-well potential has the left-right symmetry, we should investigate the odd and even parity spectra separately. 

\begin{figure}
\begin{minipage}{14pc}
\includegraphics[%
  width=0.9\linewidth,
  keepaspectratio]{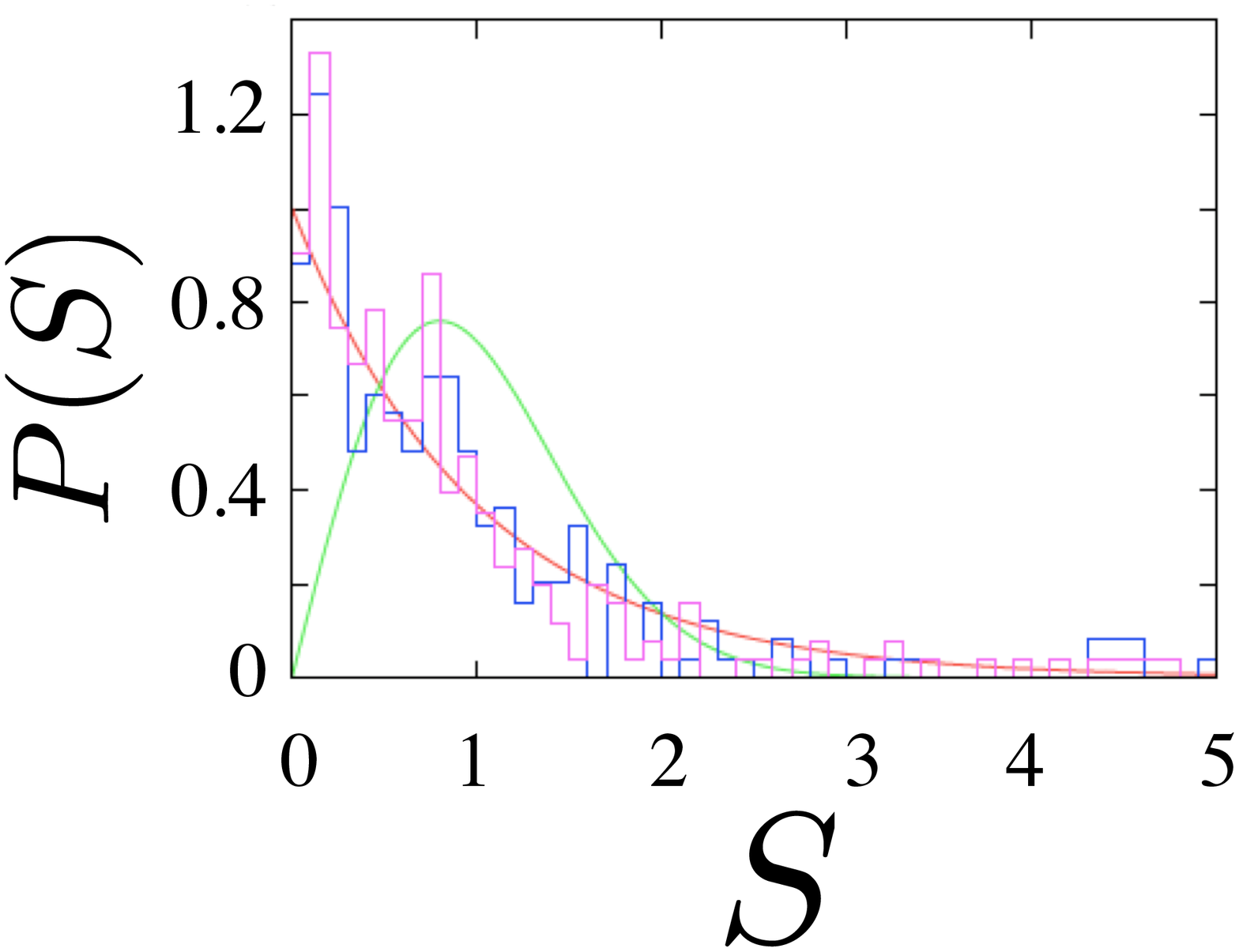}
\caption{(color online)Level spacing distributions at $\sqrt{N}g/J_{a}=0.447$. Blue and violet bars represent odd and even parity energy spectra. Red and green lines are Poisson and Wigner distributions.}
\label{LS_g_low}
\end{minipage}\hspace{0.5pc}
\begin{minipage}{14pc}
\includegraphics[%
  width=0.7\linewidth,
  keepaspectratio]{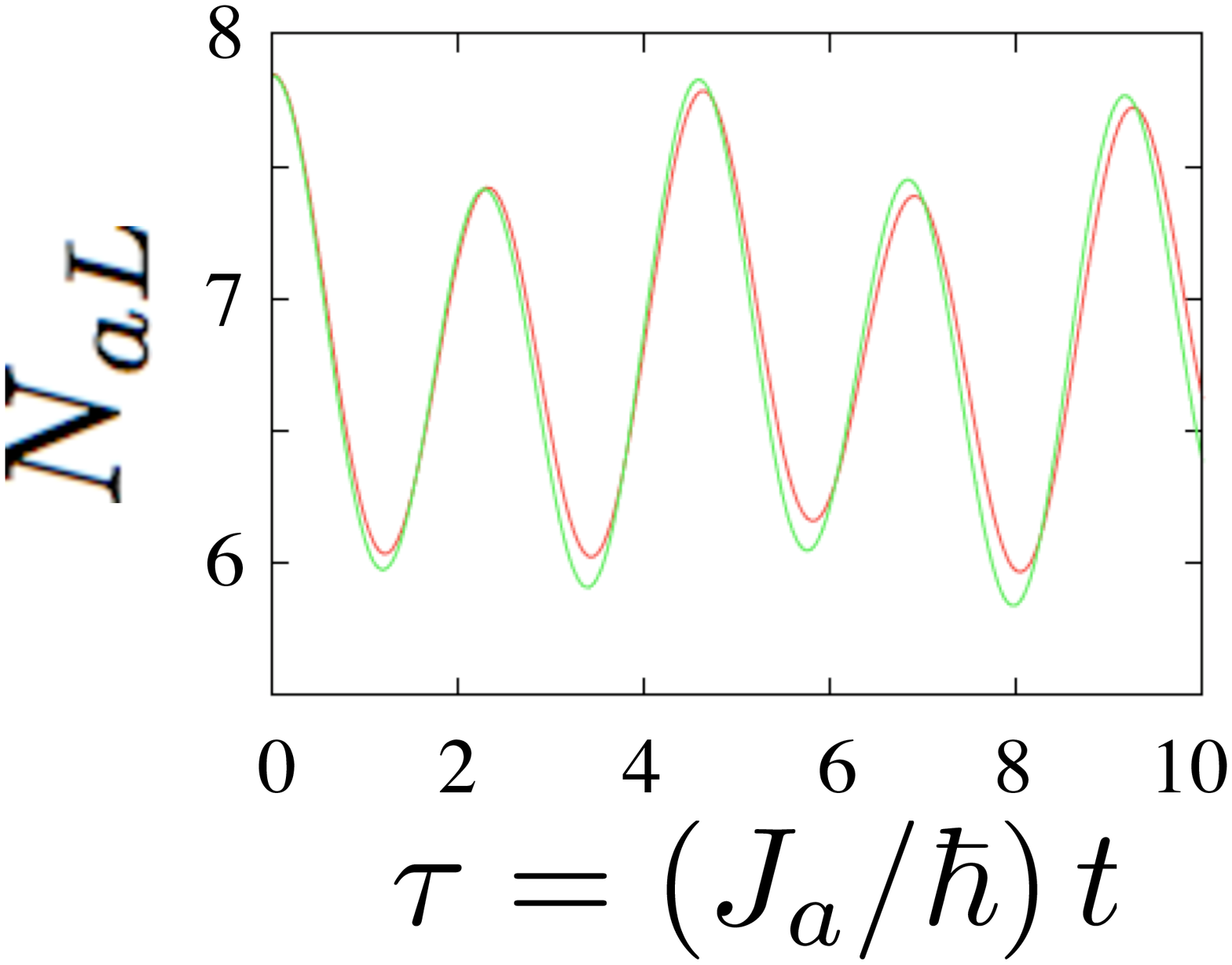}
\caption{(color online)The time evolution of $N_{aL} \equiv \langle a^{\dag}_{L} a_{L} \rangle$ at $\sqrt{N}g/J_{a}=0.447$. As the initial condition, we shift the atomic particle distribution of the ground state to the left well by a single atom. Green and red lines represent semiclassical and full-quantum time evolution.}
\label{Dynamics_g_low}
\end{minipage}
\end{figure}

\begin{figure}
\begin{minipage}{14pc}
\includegraphics[%
  width=0.9\linewidth,
  keepaspectratio]{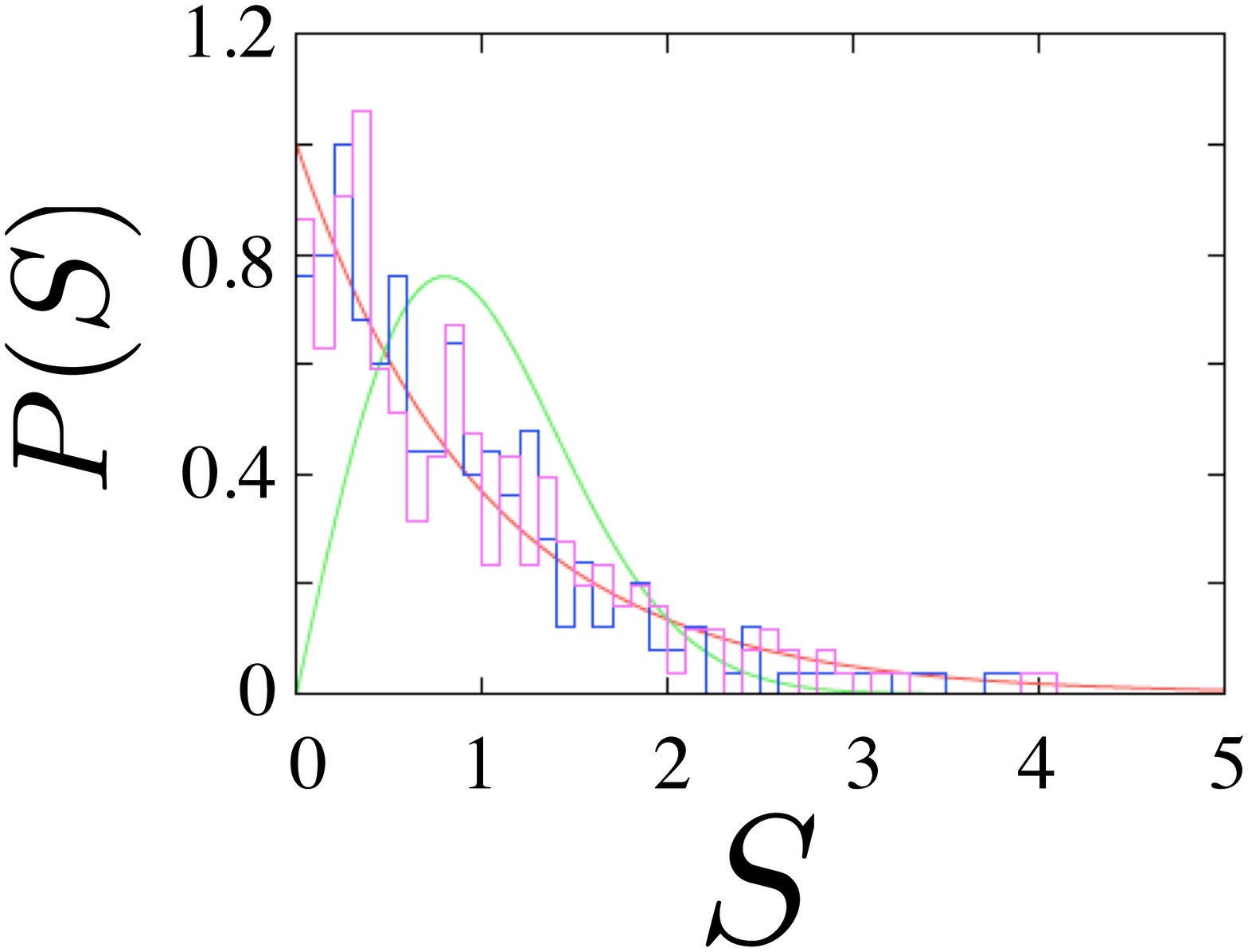}
\caption{(color online)Level spacing distributions at $\sqrt{N}g/J_{a}=33.5$. Blue and violet bars represent odd and even parity energy spectra. Red and green lines are Poisson and Wigner distributions.}
\label{LS_g_high}
\end{minipage}\hspace{0.5pc}
\begin{minipage}{14pc}
\includegraphics[%
  width=0.7\linewidth,
  keepaspectratio]{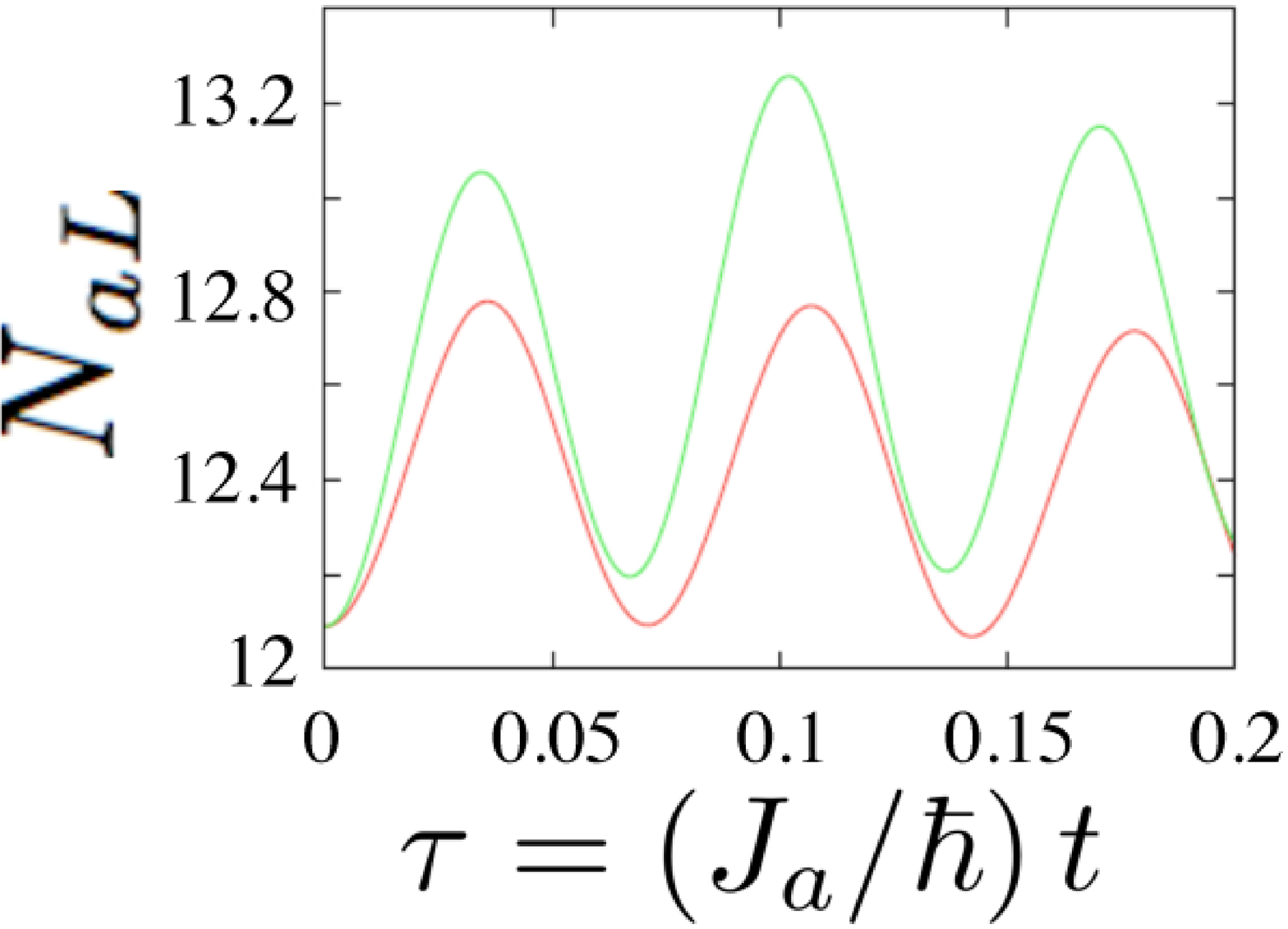}
\caption{(color online)The time evolution of $N_{aL} \equiv \langle a^{\dag}_{L} a_{L} \rangle$ at $\sqrt{N}g/J_{a}=33.5$. As the initial condition, we shift the atomic particle distribution of the ground state to the right well by a single atom. Green and red lines represent semiclassical and full-quantum time evolution.}
\label{Dynamics_g_high}
\end{minipage}
\end{figure}

From Fig. \ref{LS_g_low}, the level spacing distiribution in the small $g$ regime is very close to Poisson distribution. In this regime, Fig. \ref{Dynamics_g_low} shows that the time evolutions in a semiclassical approximation resemble the full-quantum dynamics, and both exhibit regular motions. 
Furthermore, the level spacing distribution is Poisson-like in the larger $g$ region(see Fig. \ref{LS_g_high}). The time evolutions of $N_{aL} \equiv \langle a^{\dag}_{L} a_{L} \rangle$ exhibit regular motion in both semiclassical and full-quantum treatments, as shown in Fig. \ref{Dynamics_g_high}. In contrast to Fig. \ref{Dynamics_g_low}, we see a remarkable difference between the semiclassical and full-quantum dynamics. This is due to the particle localization in one well, which decreases the particle number in another well and enhances quantum fluctuations.
In $\sqrt{N} g / J_{a} \gg 1$ and $\sqrt{ N } g / J_{a} \ll 1$ regions, internal or interwell tunneling are dominant, and thus the multi-degree of freedom of this system does not have strong influence.
\begin{figure}
\begin{minipage}{14pc}
\includegraphics[%
  width=0.9\linewidth,
  keepaspectratio]{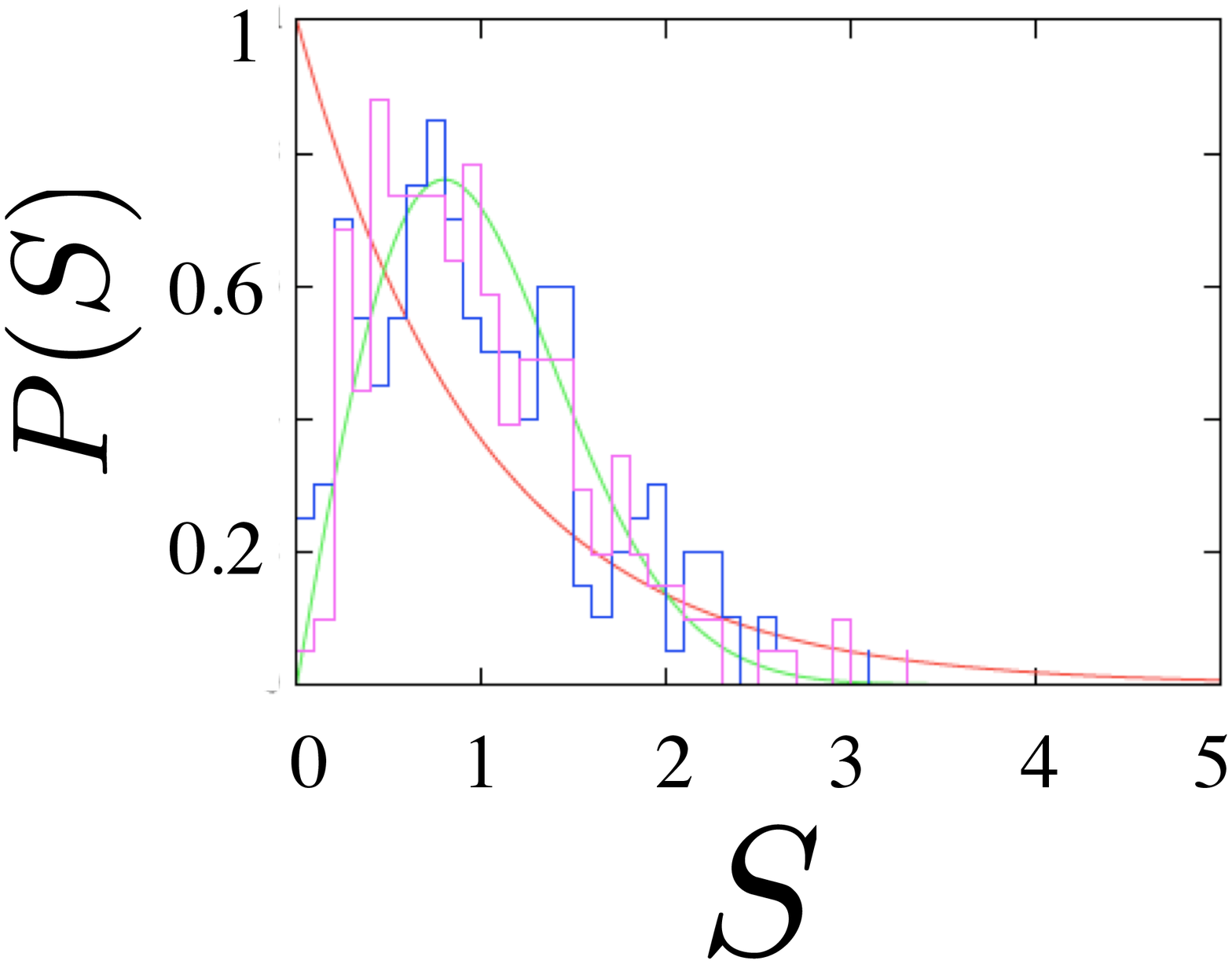}
\caption{(color online)Level spacing distributions at $\sqrt{N}g/J_{a}=3.43$. Blue and violet bars represent odd and even parity energy spectra. Red and green lines are Poisson and Wigner distributions. Lower 10 \% and higher 10 \% levels are omitted.}
\label{LS_g_mid}
\end{minipage}\hspace{0.5pc}
\begin{minipage}{14pc}
\includegraphics[%
  width=0.8\linewidth,
  keepaspectratio]{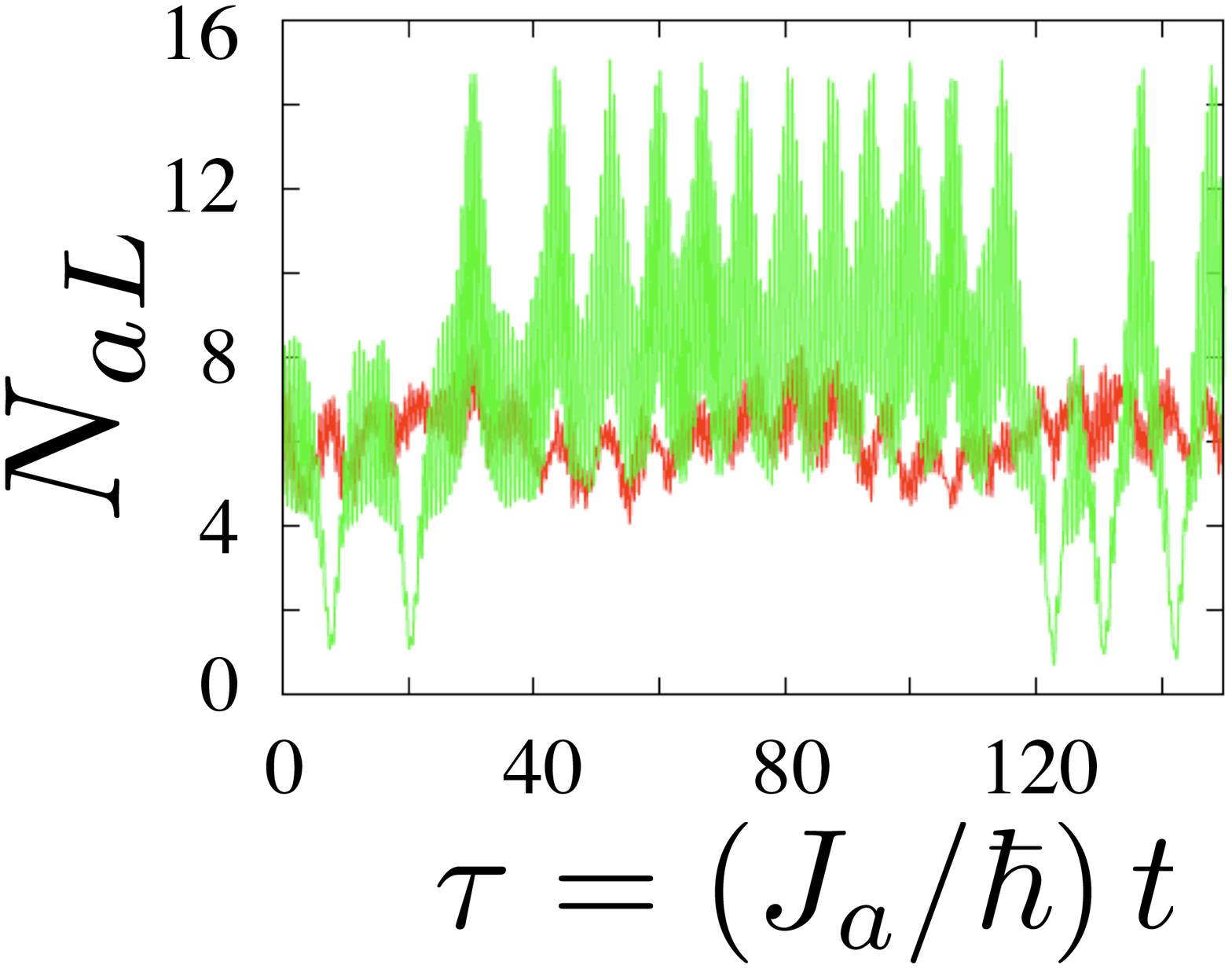}
\caption{(color online)The time evolution of $N_{aL} \equiv \langle a^{\dag}_{L} a_{L} \rangle$ at $\sqrt{N}g/J_{a}=3.43$. As the initial condition, we shift the atomic particle distribution of the ground state to the right well by five atoms. Green and red lines represent semiclassical and full-quantum time evolution.}
\label{Dynamics_g_mid}
\end{minipage}
\end{figure}
In contrast, as shown in Fig. \ref{LS_g_mid}, the almost all energy spectra obey Wigner distribution in the intermediate coupling region. Fig. \ref{Dynamics_g_mid} shows that the semiclassical dynamics is chaotic and quite different from the full-quantum time evolution. It is shown that the quantum fluctuations induce dynamical localization. The similar behavior was found in a kicked rotor\cite{Casati_Molinari}\cite{rotor}.

\section{Conclusions}
In this paper, we showed the particle localization in ground states as shown in Fig. \ref{rel_za}. Furthermore, we investigated how the internal tunneling deform the level spacing distribution and dynamics. Increasing the internal tunneling, the dynamics change to be non-periodic. Further increasing the internal tunneling, periodic time evolution emerges again. In addition, we showed that semiclassical and full-quantum dynamics is quite different in chaotic region; dynamical localization occurs within full-quantum treatment. 

\end{document}